\documentclass[onecolumn]{article}
\usepackage{amsmath}
\usepackage{amsthm}
\usepackage{float}
\usepackage{amsfonts}
\usepackage{amssymb}
\usepackage{mathrsfs}
\usepackage{graphicx}

\theoremstyle{remark}

\begin{document}


\title{Adaptive Combination of $l_0$-LMS Adaptive Filters for Sparse System Identification in Fluctuating Noise Power}

\author{Bijit K. Das, Mrityunjoy Chakraborty}

%
%

\thanks{B. K. Das and M. Chakraborty are with the Department of Electronics and Electrical Communication
Engineering, Indian Institute of Technology, Kharagpur, INDIA
(e.mail : bijitbijit@gmail.com;
 mrityun@ece.iitkgp.ernet.in).}

\abstract{Recently, the $l_0$-least mean square ($l_0$-LMS) algorithm has been proposed to identify sparse
linear systems by employing a sparsity-promoting continuous function as an approximation
of $l_0$ pseudonorm penalty. However, the performance of this algorithm is sensitive to the appropriate
choice of the some parameter responsible for the zero-attracting intensity.
The optimum choice for this parameter depends on the signal-to-noise ratio (SNR) prevailing
in the system. Thus, it becomes difficult to fix a suitable value for this parameter,
particularly in a situation where SNR fluctuates over time.
In this work, we propose several adaptive combinations of differently parameterized $l_0$-LMS
to get an overall satisfactory performance independent of the SNR, and discuss some issues relevant
to these combination structures.
We also demonstrate an efficient partial update scheme which not only reduces the number
of computations per iteration, but also achieves some interesting performance gain compared with the full update case.
Then, we propose a new recursive least squares (RLS)-type rule to update the combining parameter more efficiently.
Finally, we extend the combination of two filters to a combination of $M$ number adaptive filters,
which manifests further improvement for $M>2$.}

\maketitle


\section{Introduction}
Exploiting sparsity of the identifiable system has been a celebrated topic in the last decade among adaptive filtering
research community. However, it has a long and diverse history within and outside the adaptive signal processing community.
Researchers in the field of of network and acoustic echo cancellation have tried to take advantage of the sparse echo path models
for improving the performance of the echo cancellation algorithms. Adaptive sparse system identification has also found applications in 
the field of sparse wireless multipath channel estimation and shallow underwater acoustic communication channel estimation.
Starting from the adaptive delay filters and active tap detection based algorithms to the popular family of proportionate-type
adaptive filters and recently proposed set theoretic adaptive filters, numerous attempts have been made to use the a priori knowledge
about system sparsity.

Another family of sparsity-promoting norm regularized algorithms has gained immense popularity in recent years.
Historically, the basis pursuit and other related methods have shown advantages of this approach.
After the advent of sparse signal reconstruction techniques in the compressive sensing literatures, different 
sparsity-promoting norms have been borrowed by the adaptive filtering researchers. 

Though the $l_0$-pseudonorm which measures the sparsity by counting the number of nonzero elements in a vector 
can not be directly used for the regularization purpose since it is not a continuous function. Several 
approximations of this have been considered. The $l_1$-norm or the absolute sum, the log-sum have been studied
to a reasonable extent in this context, and have given birth to the algorithms like ZA-LMS and the RZA-LMS.
Another approximation of $l_0$-norm by some exponential function has also been proposed. This algorithm manifests 
excellent behaviour in terms of convergence speed and steady-state mean square deviation for proper choice of a
parameter responsible for zero-attracting intensity. 

In section $II$, we provide a brief review of the algorithm.

In section $III$, it has been shown that the choice of this parameter is sensitive to the signal-to-noise ratio (SNR) of the setup.
An adaptive technique has been proposed to tackle this problem but with increasing complexity of the algorithm.

In section $IV$, we propose a convex combination of two differently parameterized $l_0$-LMS adaptive filters to alleviate 
the sensitivity of selection of this parameter to some extent. We also demonstrate the simulation results to support
the advantage of the proposed scheme.

In section $V$, we discuss a reduced-complexity partial update scheme for this combination. In this section, we also derive 
a new recursive least squares (RLS) type update rule for efficient adaptation of the combining parameter. The simulation
results, we provide in this section, show how this update scheme proves to be a better alternative of the conventional
gradient descent based update rule.

In section $VI$, we extend the techniques presented in the earlier sections to a more general combination of $M$ number of 
adaptive filters.

\section{Brief Review of the $l_0$-LMS Algorithm}
For deriving the $l_0$-LMS algorithm, the cost function is modified as
\begin{eqnarray}
  L_0(n) = \frac{1}{2}e^2(n) + \gamma_{2} {\parallel {\bf w}(n) \parallel}_0
\end{eqnarray}

Considering that the $l_0$-norm minimization is a Non-Polynomial (NP) hard problem, $l_0$-norm is generally
approximated by a continuous function. A popular approximation is
\begin{eqnarray}
 {\parallel{\bf w}(n)\parallel}_0 \approx \sum\limits_{i=1}^{L} \left(1 - e^{-\beta \arrowvert w_{i}(n) \arrowvert } \right) 
\end{eqnarray}

By minimizing the above cost function the gradient descent recursion for the $i^{th}$ filter coefficient becomes
\begin{eqnarray}
 & & w_{i}(n+1) = w_i(n) - \kappa \beta sgn\{ w_i(n) \}e^{-\beta \arrowvert w_{i}(n) \arrowvert} \nonumber\\
 & & + \mu e(n)x(n-i+1) \nonumber\\
 & & (\forall 1 \leq i \leq L) 
\end{eqnarray}

To reduce the computational complexity, the first order Taylor series expansion of exponential functions is taken into consideration, 
\begin{eqnarray}
 e^{-\beta \arrowvert a \arrowvert} &\approx& 1 - \beta \arrowvert a \arrowvert \hspace{4mm} \text{if} \hspace{2mm} \arrowvert a \arrowvert < \cfrac{1}{\beta} \nonumber\\
 & \approx & 0 \hspace{4mm} \text{otherwise} 
\end{eqnarray}
It is to be noted that the above approximation is bounded to be positive because the exponential function is larger than zero.\\

Now, the approximated gradient descent recursion is
\begin{eqnarray}
 & & w_{i}(n+1) = w_i(n) - \kappa \beta f_{\beta}(w_i(n)) + \mu e(n)x(n-i+1),  \nonumber\\
 & & (\forall 1 \leq i \leq L), 
\end{eqnarray}
where
\begin{eqnarray}
f_{\beta}(a) & = & \beta^2 a + \beta, \hspace{4mm} \text{if} \hspace{2mm} -\cfrac{1}{\beta} \leq a \leq 0;\nonumber\\
             & = & \beta^2 a - \beta, \hspace{4mm} \text{if} \hspace{2mm} 0 \leq a \leq \cfrac{1}{\beta};\nonumber\\
             & = & 0 \hspace{4mm} \text{elsewhere}\nonumber
\end{eqnarray}

\section{The Dependence of the Optimum $\kappa$ on the Signal-to-Noise Ratio (SNR)}
In this section, we try to show how the optimum $\kappa$ depends on the signal-to-noise ratio (SNR).
\begin{figure*}[h]
\begin{center}
\includegraphics[width=160mm, height=55mm]{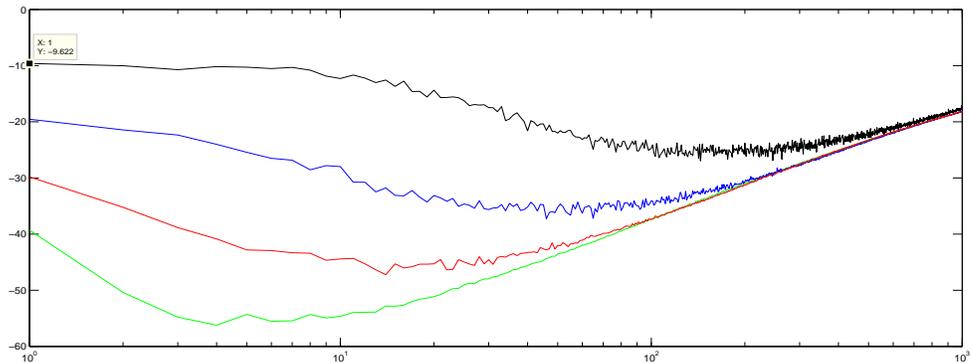}
\end{center}
\caption{Steady-state MSD vs. $\kappa$ for four SNR values [$10$dB (black), $20$dB (blue), $30$dB (red), $40$dB (green)])}.
\vspace*{-3pt}
\end{figure*}\\\\

\emph{Experiment $1$:}
We try to identify a sparse FIR system of length $128$, which has only five nonzero coefficients (set as $1$), and rest of them are zeros.
We use a zero mean, unit variance white Gaussian random process as input to the unknown system as well as the $l_0$-LMS adaptive filter.
The parameters of the algorithm are set as: $\mu=0.8$, . Matlab simulations are run for $15000$ samples and steady-state MSD is calculated
by averaging repeating the experiment over $100$ iterations. The value of $\kappa$ is varied from $0.000001$ to $0.0001$ with $1000$ steps, 
and the steady-state MSD is plotted against $\kappa$ in Fig.$1$. Four different noise variances are chosen for the experiment. The black, blue, red and
green curves correspond to $10$dB, $20$dB, $30$dB, $40$dB SNR-s respectively.

The optimum $\kappa$ for which the minimum MSD is achieved is different for different SNRs. It explicitly implies that when SNR is likely to
fluctuate over a wide range, a fixed single $\kappa$ can not guarantee the best result for all different cases. This fact motivates us to propose a convex combination
of two $l_0$-LMS adaptive filters with two different $\kappa$ values.

\section{Proposed Adaptive Convex Combination of Two Differently Parameterized $l_0$-LMS Adaptive Filters}
In this section, we propose a convex combination of two $l_0$-LMS adaptive filters. In this part of the paper, we have constrained
ourselves to use the convex combination scheme in [].
The outputs of the two $l_0$-LMS adaptive filters are combined using the following rule:
\begin{eqnarray}
 y(n) = \lambda(n)y_1(n) + (1-\lambda(n))y_2(n)
\end{eqnarray}
where $y_i(n)$ is the output of the $i^{th}$ adaptive filter ($i=1,2$), and $y(n)$ is the combined output.

These two filters are identical in all aspects apart from their individual values of $\kappa$.
\begin{eqnarray}
 \lambda(n)=\cfrac{1}{1+exp(-a(n))}
\end{eqnarray}

The update rule for the parameter $a(n)$ is as follows:
\begin{eqnarray}
 a(n+1) = a(n) + \mu_c e(n)(y_1(n)-y_2(n))\lambda(n)(1-\lambda(n)),
\end{eqnarray}
where $e(n) = d(n) - y(n)$.

It is also noteworthy that $a(n)$ is constrained within the range of $\pm5$. This is a common practice to avoid
$a(n)$ being stucked at any of the two extremes of its unconstrained range.

\emph{Experiment $2$:}
The $\kappa$ values are chosen as $5\times 10^{-5} $ and $5\times 10^{-6} $.
Three SNR values are chosen as $60$dB, $40$dB and $20$dB. The simulation starts with $60$dB SNR, at the $6000^{th}$ iteration noise
power increases to give a SNR of $40$dB, and finally, it becomes $20$dB at the $12000^{th}$ iteration. The simulation stops at the 
$18000^{th}$ iteration. The MSD-s are obtained by averaging $100$ independent runs of this simulation. The $mu_c$ is kept at $3000$.
All other parameters and variables are same as in the experiment $1$.
The instantaneous MSD-s are plotted against iteration index $n$ in fig.2.
The proposed combination (plotted in green) achieves the best steady-state MSDs for all SNR levels. The red and the black curves
represent two $l_0$-LMS filters (for $\kappa=5\times 10^{-5}$ and $5\times 10^{-6} $ respectively.)

\begin{figure*}[h]
\begin{center}
\includegraphics[width=160mm, height=65mm]{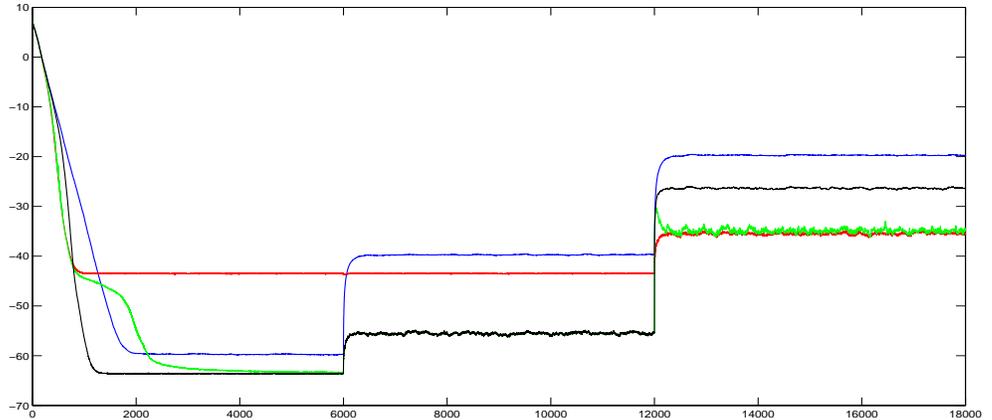}
\end{center}
\caption{Instantaneous MSD Curves for the Proposed Combination (Green), $l_0$-LMS (with $\kappa=5\times 10^{-5} $)(Red), $l_0$-LMS (with $\kappa= 5\times 10^{-6}$)(Black),
LMS (Blue)}.
\vspace*{-3pt}
\end{figure*}

\section{A Reduced Complexity Partial Update Scheme}
One of the major drawbacks of the combination schemes, in general, is the increase in the number of computations per iteration by a multiple factor
for deployment of multiple adaptive filters simultaneously. In this section, we demonstrate a partial update scheme for reducing the computational burden.
We divide the set of all filter taps into two mutually exclusive subsets, named as $Odd$ and $Even$ for all odd and all even taps respectively. 
We update the filter taps of these individual subsets at alternative iterations.
Now, one natural question that can arise is how we choose the subsets for individual component filters. Is it same for all the filters?
Or, it is rather more beneficial to choose different subsets for different filters at a particular iteration. For the time being,
we go for the second option without discussing the relative merits and demerits of this choice. We leave this discussion for the section $VI$.

\begin{table}[ht]
\caption{The Partial Update Scheme with Mutually Exclusive Subsets for $M$ Number of Filters} \fbox{
 \addtolength{\linewidth}{-2\fboxsep}%
 \addtolength{\linewidth}{-2\fboxrule}%
 \begin{minipage}[t][1.2\height][c]{0.9\linewidth}
 Initialization : ${\bf w}_{k}(0) = 0$ for each adaptive filter $k$.
 $M$ number of diagonal matrices defined as ${\bf S}_i \hspace{3mm} (i=1,\cdots,M)$, for which $[{\bf S}_i]_{j,j} = 1$ for $j = i, i+M, i+2M, \cdots, i+\left\lfloor \cfrac{L}{M} \right\rfloor M$, and $= 0$ otherwise.\\
For every index $n$ and every adaptive filter $k$, repeat
  \begin{eqnarray}
     e_k(n) &=& d_k(n) - {\bf w}_{k}^{T}(n){\bf x}(n)\nonumber\\
     {\bf P}_{k}(n) &=& {\bf S}_{l}\nonumber\\
    & & (l = quotient(k+n, M))   \nonumber\\                  
    {\bf w}_k(n+1) &=& {\bf w}_k(n) +  {\bf P}_k(n)[\mu_k {\bf x}_k(n)e_k(n) + \kappa_k \beta f_{\beta}(w_k(n))]
  \end{eqnarray}
  \end{minipage}
}
  \label{table:PU_me}
\end{table}

\subsection{A New RLS-type Update Rule for Adapting the Combining Parameter}
We also derive a new update rule for $a(n)$ in the following.
Let us first define the least-squares cost function $J_c(n)$ as
\begin{eqnarray}
 J_c(n) = \sum_{k=0}^{n}\beta^{k}e_{c}^{2}(n-k),
\end{eqnarray}
where $ e_c(n) = d(n) - y_c(n)$ is the error at the $n^{th}$ instant for the combined filter.

$J_c(n)$ can be expanded by 
\begin{eqnarray}
 J_c(n) &=& \sum_{k=0}^{n}\beta^{k}[d(n-k)-\lambda(n)\{y_1(n-k)-y_2(n-k)\} + y_2(n-k)]^2
\end{eqnarray}
The optimum value of $\lambda(n)$, $\lambda_{opt}(n)$ can be found as
\begin{eqnarray}
 & &\lambda_{opt}(n) = [\sum_{k=0}^{n}\beta^{k}\{y_1(n-k)-y_2(n-k)\}^{2}]^{-1}[\sum_{m=0}^{n}\beta^{m}\{y_1(n-m)-y_2(n-m)\}\{d(n-m) - y_2(n-m) \}]\nonumber\\
 &=& [ {\bf y}_{d}^{T}(n){\bf B}(n){\bf y}_{d}(n)]^{-1} [{\bf y}_{d}^{T}(n){\bf B}(n){\bf p}(n)],\\
 \end{eqnarray}
where
 $ p(n) = d(n) - y_2(n)$, and ${\bf p}(n) = [p(n), p(n-1), \cdots, p(0)]^{T} $. \\\\

 Now,defining $r_{in}(n) = \left[{\bf y}_{d}^{T}(n){\bf B}(n){\bf y}_{d}(n)\right]^{-1}$,
 $ y_d(n) = y_1(n) - y_2(n)$,
 ${\bf y}_d(n) = [y_d(n), y_d(n-1), \cdots, y_d(0)]^{T} $, we get
 \begin{eqnarray}
 & &\lambda_{opt}(n) = r_{in}(n)[{\bf y}_{d}^{T}(n){\bf B}(n){\bf p}(n)]
 \end{eqnarray}

Now, to derive a recursive update for $\lambda_{opt}(n)$ using $y_d(n+1)$ and $p(n+1)$, we write
\begin{eqnarray}
& & \lambda_{opt}(n+1)\nonumber\\
&=&  r_{in}(n+1)[{\bf y}_{d}^{T}(n+1){\bf B}(n+1){\bf p}(n+1)]\nonumber\\
 &=& [ \beta r_{in}^{-1}(n) + y_{d}^{2}(n+1)]^{-1} \{ \beta {\bf y}_{d}^{T}(n){\bf B}(n){\bf p}(n) + y_{d}(n+1)p(n+1)\}  \nonumber\\
 &=& \left\{\cfrac{1}{\beta} r_{in}(n) - \cfrac{1}{\beta^2} y_{d}^{2}(n+1)r_{in}(n)\left[1 + y_{d}^{2}(n+1)r_{in}(n)\right]^{-1}r_{in}(n)\right\} \nonumber\\
 & & \left \{ \beta {\bf y}_{d}^{T}(n){\bf B}(n){\bf p}(n) + y_{d}(n+1)p(n+1)\right\}  \nonumber\\
 &=& \lambda_{opt}(n) -  \cfrac{1}{\beta} y_{d}^{2}(n+1)\left[1 + \cfrac{1}{\beta}y_{d}^{2}(n+1)r_{in}(n)\right]^{-1}r_{in}(n)\lambda_{opt}(n) +  \cfrac{1}{\beta} r_{in}(n)\nonumber\\
 & & \left[1 - \cfrac{1}{\beta}y_{d}^{2}(n+1)[1 + \cfrac{1}{\beta}y_{d}^{2}(n+1)r_{in}(n)]^{-1}r_{in}(n) \right]y_{d}(n+1)p(n+1) \nonumber\\
 &=& \lambda_{opt}(n) -  \cfrac{1}{\beta} y_{d}^{2}(n+1)\left[1 + \cfrac{1}{\beta}y_{d}^{2}(n+1)r_{in}(n)\right]^{-1} r_{in}(n)\lambda_{opt}(n) +  \cfrac{1}{\beta} r_{in}(n)\nonumber\\
 & & \left[1 + \cfrac{1}{\beta}y_{d}^{2}(n+1)r_{in}(n)\right]^{-1}y_{d}(n+1)p(n+1) \nonumber\\
 &=& \lambda_{opt}(n) + \cfrac{1}{\beta} r_{in}(n) \left[1 + \cfrac{1}{\beta}y_{d}^{2}(n+1)r_{in}(n)\right]^{-1} y_{d}(n+1)\left[p(n) - \lambda_{opt}(n)y_d(n) \right] \nonumber\\
 &=& \lambda_{opt}(n) + \cfrac{1}{\beta} r_{in}(n) \left[1 + \cfrac{1}{\beta}y_{d}^{2}(n+1)r_{in}(n)\right]^{-1} y_{d}(n+1)\left[d(n) - \lambda_{opt}(n)y_1(n) - (1-\lambda_{opt}(n))y_2(n) \right] \nonumber\\
  &=& \lambda_{opt}(n) + \cfrac{1}{\beta} r_{in}(n) \left[1 + \cfrac{1}{\beta}y_{d}^{2}(n+1)r_{in}(n)\right]^{-1} y_{d}(n+1)e_c(n) \nonumber\\
  &=& \lambda_{opt}(n) + k(n)e_c(n),
 \end{eqnarray}
 where \begin{eqnarray}
        k(n) = \cfrac{1}{\beta} r_{in}(n) \left[1 + \cfrac{1}{\beta}y_{d}^{2}(n+1)r_{in}(n)\right]^{-1} y_{d}(n+1)
       \end{eqnarray}\\
       
       But, since this update does not constrain $\lambda_{opt}(n)$ within $0$ and $1$, we update the corresponding variable $a_{opt}(n)$, and then compute
       $\lambda_{opt}(n)$ using the aforementioned rule. Since $\lambda_{opt}(n)$ is a monotonically increasing function $a_{opt}(n)$, the same incremental update has been used
       for $a_{opt}(n)$ as above.
 \begin{eqnarray}
   a_{opt}(n+1) = a_{opt}(n) + k(n)e_c(n).
 \end{eqnarray}
 
 \emph{Experiment $3$}:
 Now, we repeat the experiment $2$ with both the conventional gradient descent based update and the propose RLS-type update for the combiner separately, and
 we also incorporate the aforementioned partial update with mutually exclusive subsets scheme mentioned in the table I for $M=2$ case. We plot both the MSDs as well as the individual MSDs for 
 the component filters and the simple LMS adaptive filter in fig.4 and fig.5 for SNR $20$dB and $40$dB respectively.
 For gradient descent based update, we plot two different curves for two different values of $\mu_c$, i.e., $1000$ and $10000$ .
 For a particular SNR, one proves to be better than other. But, the RLS-type update shows its superiority by providing satisfactory behaviour in both
 situations without tuning its forgetting factor.

\begin{figure*}[h]
\begin{center}
\includegraphics[width=160mm, height=65mm]{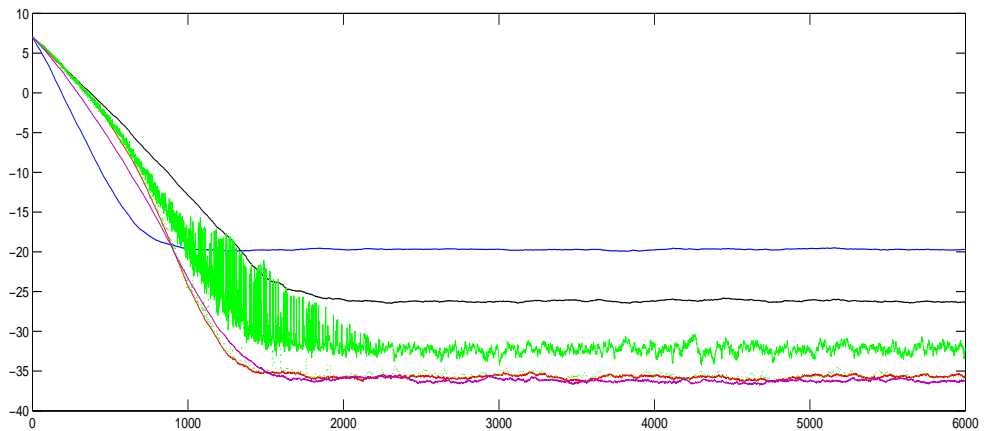}
\end{center}
\caption{Instantaneous MSD Curves for the Combination (Green), $l_0$-LMS (with $\kappa=5\times 10^{-5}$)(Red), $l_0$-LMS (with $\kappa=5\times 10^{-6}$)(Black), LMS (Blue)}.
\vspace*{-3pt}
\end{figure*}

\begin{figure*}[h]
\begin{center}
\includegraphics[width=160mm, height=65mm]{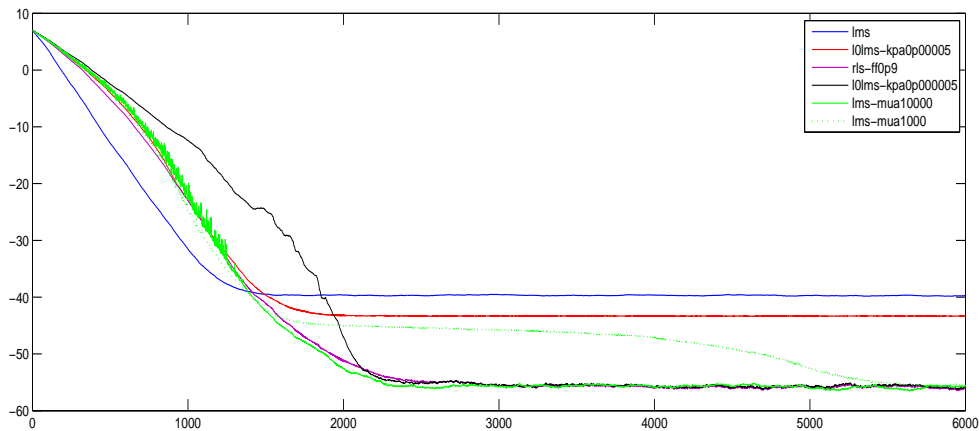}
\end{center}
\caption{Instantaneous MSD Curves for the Proposed Combination (Green), $l_0$-LMS (with $\kappa= $)(Red), $l_0$-LMS (with $\kappa= $)(Black), LMS (Blue)}.
\vspace*{-3pt}
\end{figure*}

\section{An Adaptive Convex Combination of $M$ Number of Filters}

In this section, we extend our work to a more general case of combining $M$ adaptive filters so that a much wider range of SNR
can be covered.
We also modify the RLS-type combiner update rule for this general case. 
The derivation is skipped here.
The rule is shown in the Table $2$.

\begin{table}[ht]
\caption{The Partial Update Scheme with Mutually Exclusive Subsets for $M$ Number of Filters} \fbox{
 \addtolength{\linewidth}{-2\fboxsep}%
 \addtolength{\linewidth}{-2\fboxrule}%
 \begin{minipage}[t][1.2\height][c]{0.9\linewidth}
 Initialization : ${\bf w}_{k}(0) = 0$ for each $k$.\\
 ${\bf S}_l[j,j] = 1$ for $j=l:\left\lfloor \cfrac{N}{M}\right\rfloor:N$\\
 and $= 0$ elsewhere. \\
 (for $l=1$ to $M$)\\
For every index $n$ and every $k^{th}$ adaptive filter, repeat
  \begin{eqnarray}
      y_k(n) &=& {\bf w}_{k}^{T}(n){\bf u}(n)\nonumber\\
      e_k(n) &=& d_k(n) - y_k(n)\nonumber\\
      {\bf w}_k(n+1) &=& {\bf w}_k(n) + {\bf S}_{mod(n+k,M)}[\mu_k{\bf u}(n)e_k(n) + \kappa_k\beta f_{\beta}({\bf w}_k(n))]\nonumber\\
      \psi_k(n) &=& \cfrac{exp(\phi_k(n))}{\sum_{t=1}^{M}exp(\phi_t(n))} \nonumber\\
      y_c(n) &=& \sum_{k=1}^{M}{\psi}_k(n)y_k(n)\nonumber\\
      k_k(n) &=& \cfrac{\lambda_{f}^{-1}p_k(n)(y_c(n)-y_k(n))}{1+\lambda_{f}^{-1}p_k(n)(y_c(n)-y_k(n))^2}\nonumber\\ 
      \phi_{k}(n+1) &=& \phi_k(n) + k_k(n)e_k(n)\nonumber\\
      p_k(n+1) &=& \cfrac{1}{\lambda_{f}p_{k}^{-1}(n)+(y_c(n)-y_k(n))^2} \nonumber\\
  \end{eqnarray}

  \end{minipage}
}
  \label{table:M_rls_pu}
\end{table}

\emph{Experiment $4$}:
We perform four separate experiments with three different SNR levels, i.e. $20$dB, $40$dB, $60$dB and $40$dB (with a different unknown system described below).
In each case, the new combined adaptive filter ($M=4$) is deployed to identify the aforementioned sparse unknown FIR system. The parameters of the filters remain same except $\kappa$ chosen as $0 $, $1\times 10^{-6} $, $ 1 \times 10^{-5}$ and $5 \times 10^{-5} $.
In each case, the robustness of the combination filter (in magenta) can be observed. In fig.$8$, a particular case
has been studied where the unknown system is chosen as a near-sparse
FIR system with all zero coefficients of the unknown system of the earlier experiments are replaced with very small non-zero values.

\begin{figure*}[h]
\begin{center}
\includegraphics[width=160mm, height=65mm]{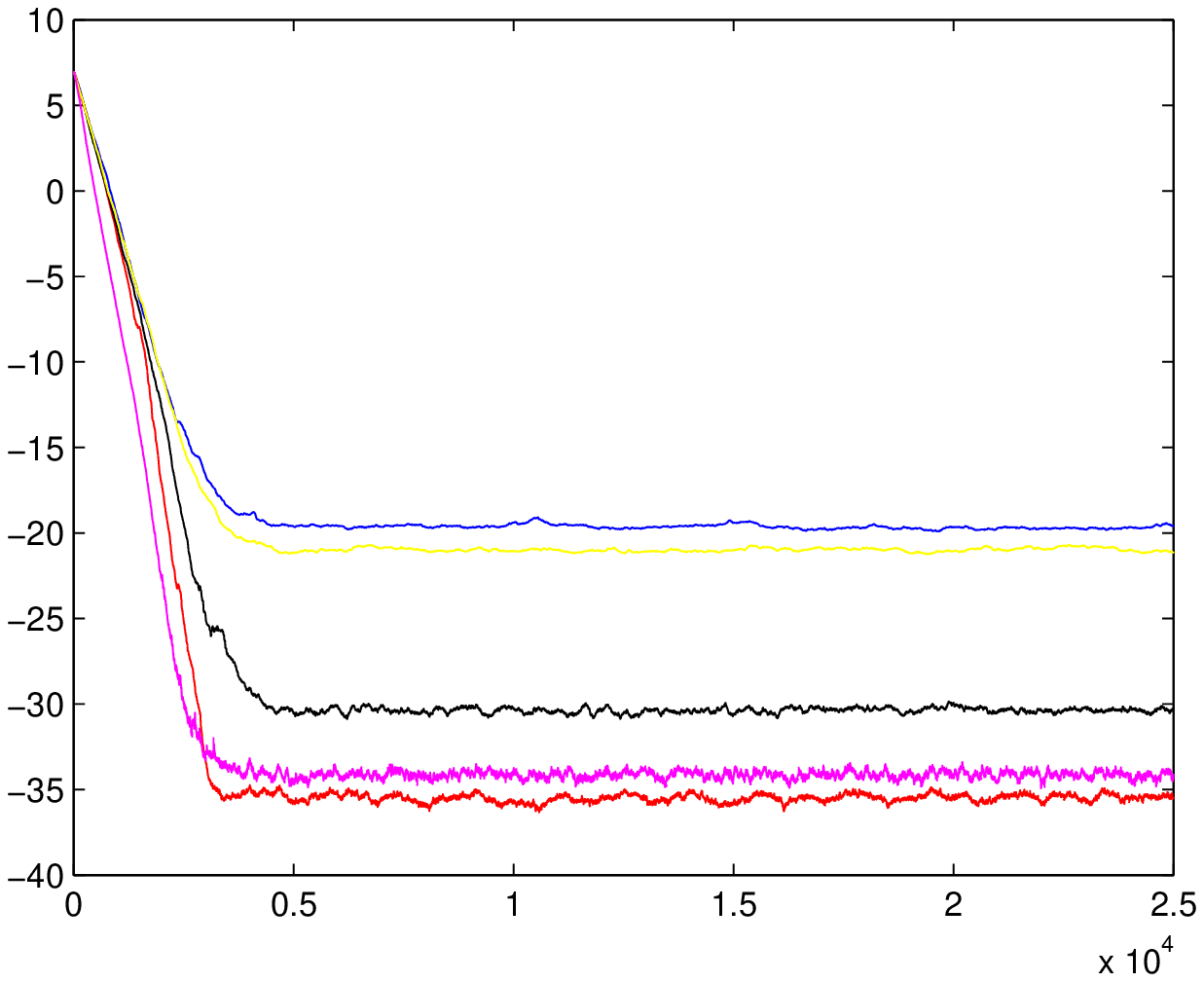}
\end{center}
\caption{MSD vs. Iteration Index for Combination of $M=4$ Adaptive Filters for SNR = $20$dB}.
\vspace*{-3pt}
\end{figure*}

\begin{figure*}[h]
\begin{center}
\includegraphics[width=160mm, height=65mm]{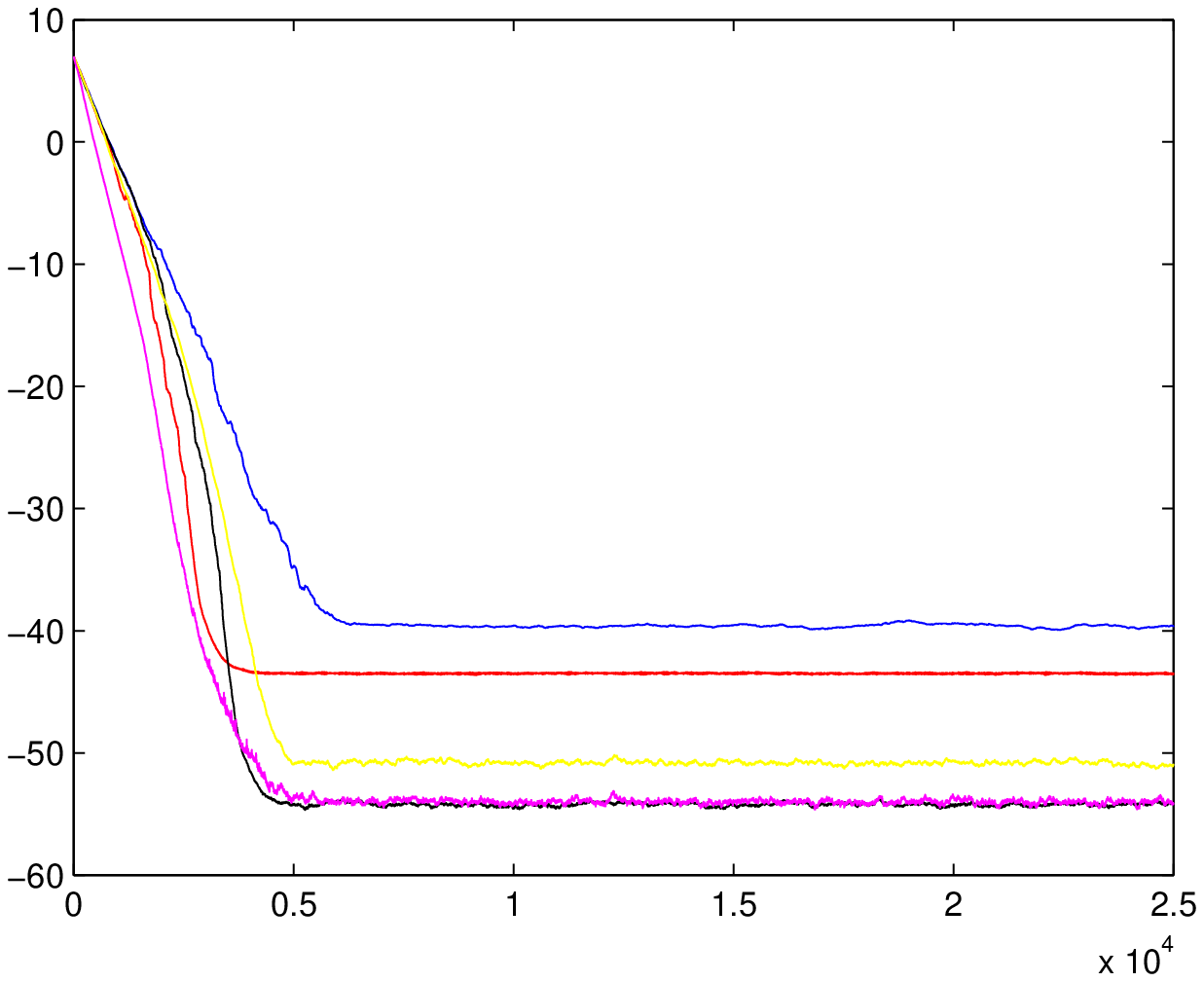}
\end{center}
\caption{MSD vs. Iteration Index for Combination of $M=4$ Adaptive Filters for SNR = $40$dB}.
\vspace*{-3pt}
\end{figure*}

\begin{figure*}[h]
\begin{center}
\includegraphics[width=160mm, height=65mm]{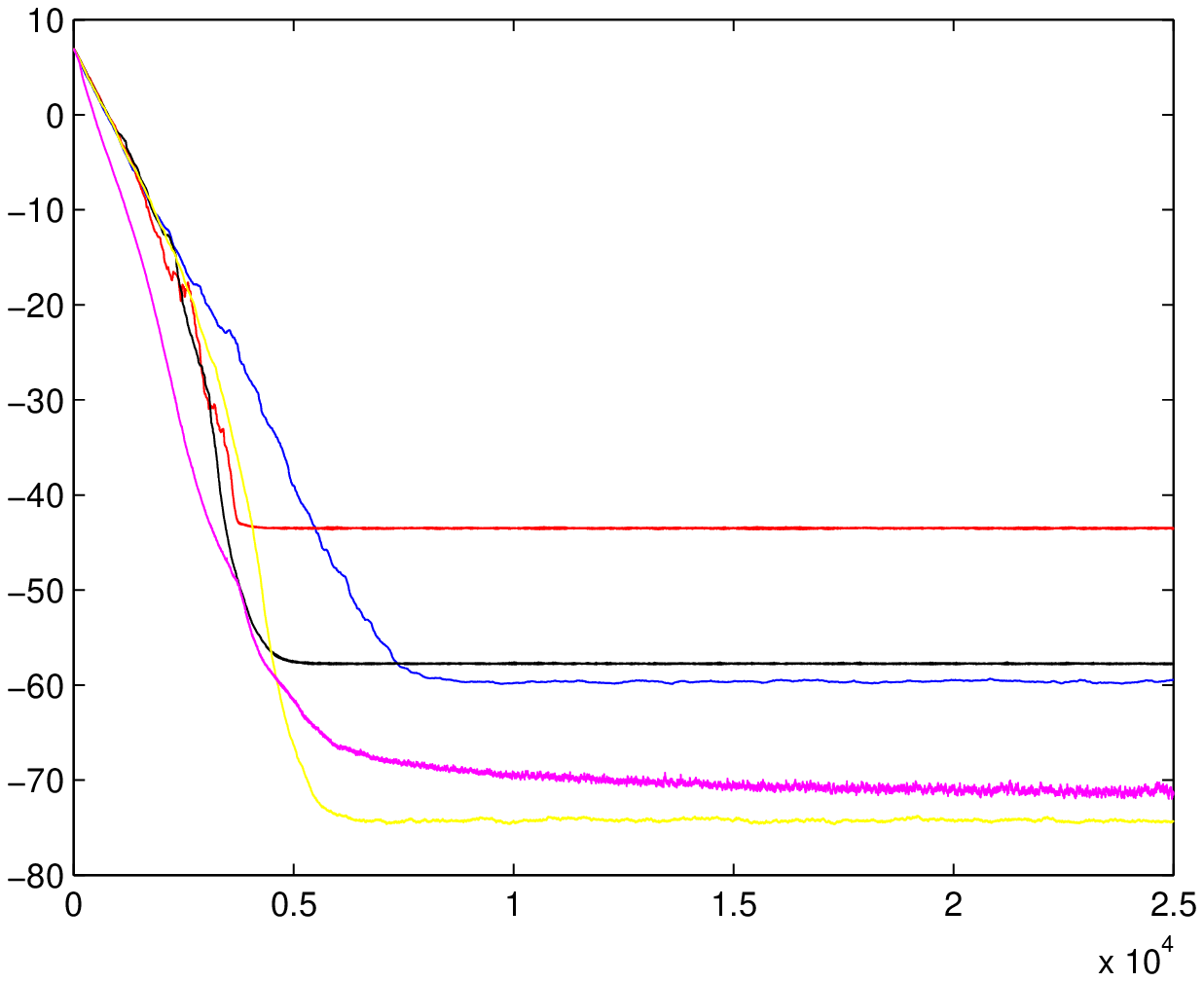}
\end{center}
\caption{MSD vs. Iteration Index for Combination of $M=4$ Adaptive Filters for SNR = $60$dB}.
\vspace*{-3pt}
\end{figure*}

\begin{figure*}[h]
\begin{center}
\includegraphics[width=160mm, height=65mm]{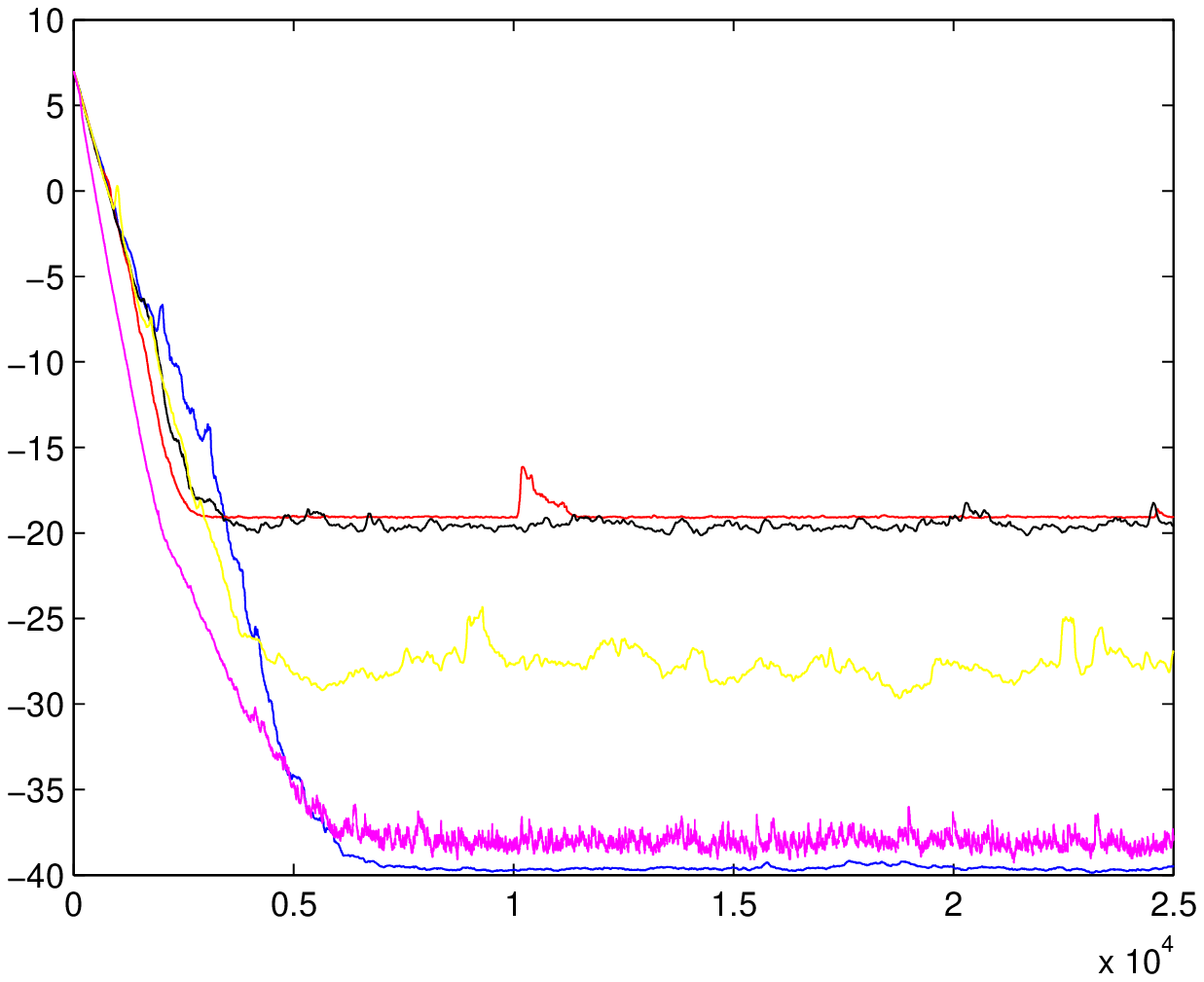}
\end{center}
\caption{MSD vs. Iteration Index for Combination of $M=4$ Adaptive Filters for SNR = $40$dB with near-sparse unknown system}.
\vspace*{-3pt}
\end{figure*}

\subsection{A Special Advantage of Using Partial Update with Mutually Exclusive Subsets}

The next three figures demonstrate a special advantage of using the partial update scheme with mutually exclusive
subsets compared with full update and PU with non-exclusive subsets. SNR level is $20$dB and the unknown system is the near-sparse one in the last experiment.
Here, though none of the component filters shows substantially better than the LMS, the combination manifests excellent performance for
the PU with mutually exclusive subsets scheme [Fig.$9$]. Fig.$10$ and fig.$11$ are respectively for the full update and the PU with same subsets.

\begin{figure*}[h]
\begin{center}
\includegraphics[width=160mm, height=65mm]{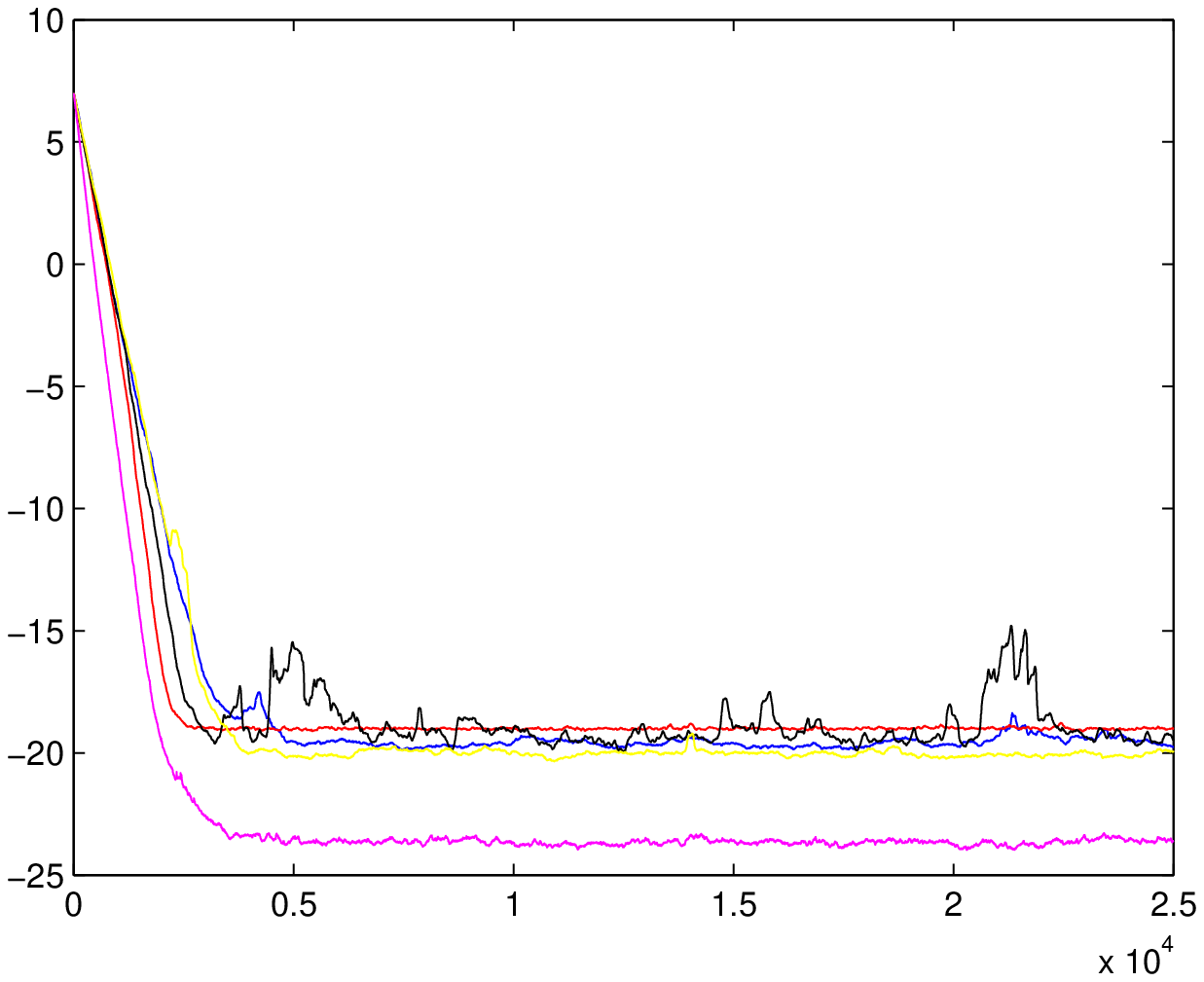}
\end{center}
\caption{MSD vs. Iteration Index for Combination (PU with mutually exclusive subsets) of $M=4$ Adaptive Filters for SNR = $20$dB with near-sparse unknown system}.
\vspace*{-3pt}
\end{figure*}

\begin{figure*}[h]
\begin{center}
\includegraphics[width=160mm, height=65mm]{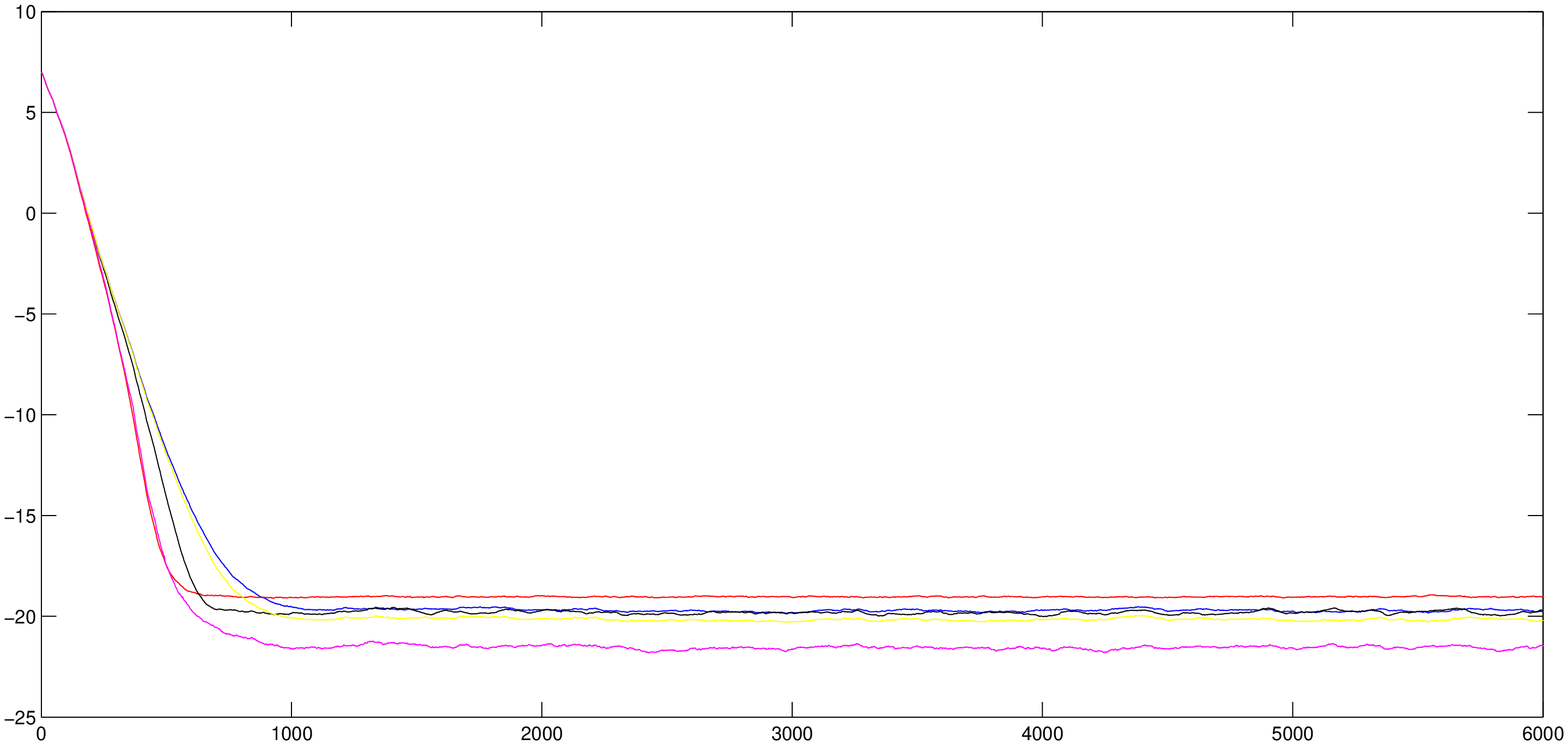}
\end{center}
\caption{MSD vs. Iteration Index for Combination (full update) of $M=4$ Adaptive Filters for SNR = $20$dB with near-sparse unknown system}.
\vspace*{-3pt}
\end{figure*}

\begin{figure*}[h]
\begin{center}
\includegraphics[width=160mm, height=65mm]{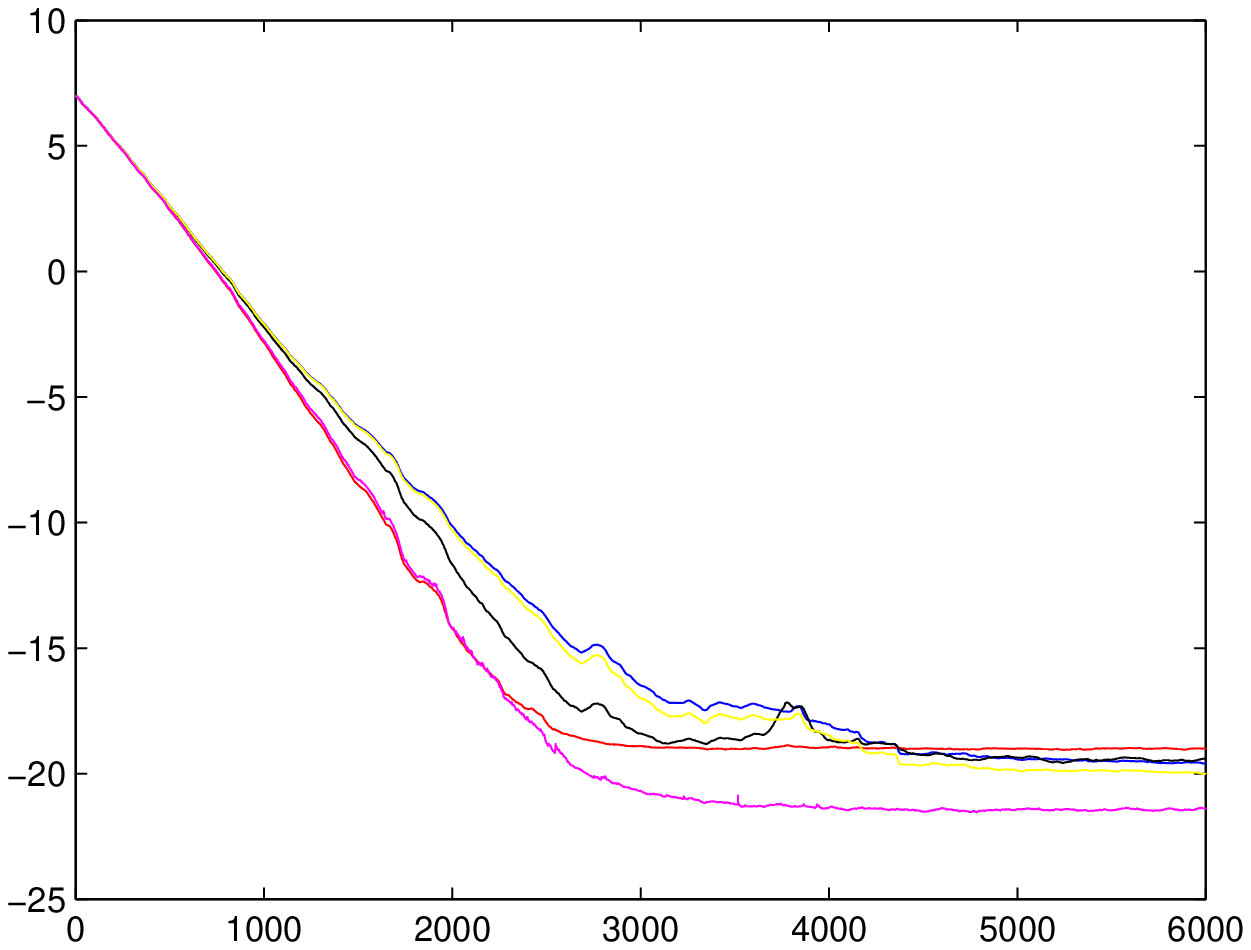}
\end{center}
\caption{MSD vs. Iteration Index for Combination (PU with same subsets) of $M=4$ Adaptive Filters for SNR = $20$dB with near-sparse unknown system}.
\vspace*{-3pt}
\end{figure*}

\section{Partial Update with Uneven Mutually Exclusive Subsets}

One major disadvantage of the partial update schemes mentioned in the last section is that it slows down the convergence 
of the individual adaptive filters as well as the combination by a fraction of $M$. In this section, we describe an uneven 
division of the filter tap indices into the subsets, and how it helps to alleviate the problem of slow convergence to some extent.

Let us here describe the uneven division of the subsets ${\bf S}_l$ and how it differs from the one described in the table II. The rule is as follows:\\

${\bf S}_l[j,j] = 1$ for $j=l:\vartriangle l:N$\\
 and $= 0$ elsewhere. \\
 (for $l=1$ to $M$. $\vartriangle l$ is the smallest integer power of $2$ greater or equal to $l$)\\

The presence of subsets with larger cardinalities help corresponding individual adaptive filters converge faster than those with smaller subsets. Thus, it leads to 
a better convergence behaviour for the overall combination as it is seen in the following figures [Fig. 10 -13]. The magenta curve (the combination) always follows
the blue one (one with lagest subset) first, then reconverges to the one with lesser steady-state m.s.d.

\begin{figure*}[h]
\begin{center}
\includegraphics[width=160mm, height=65mm]{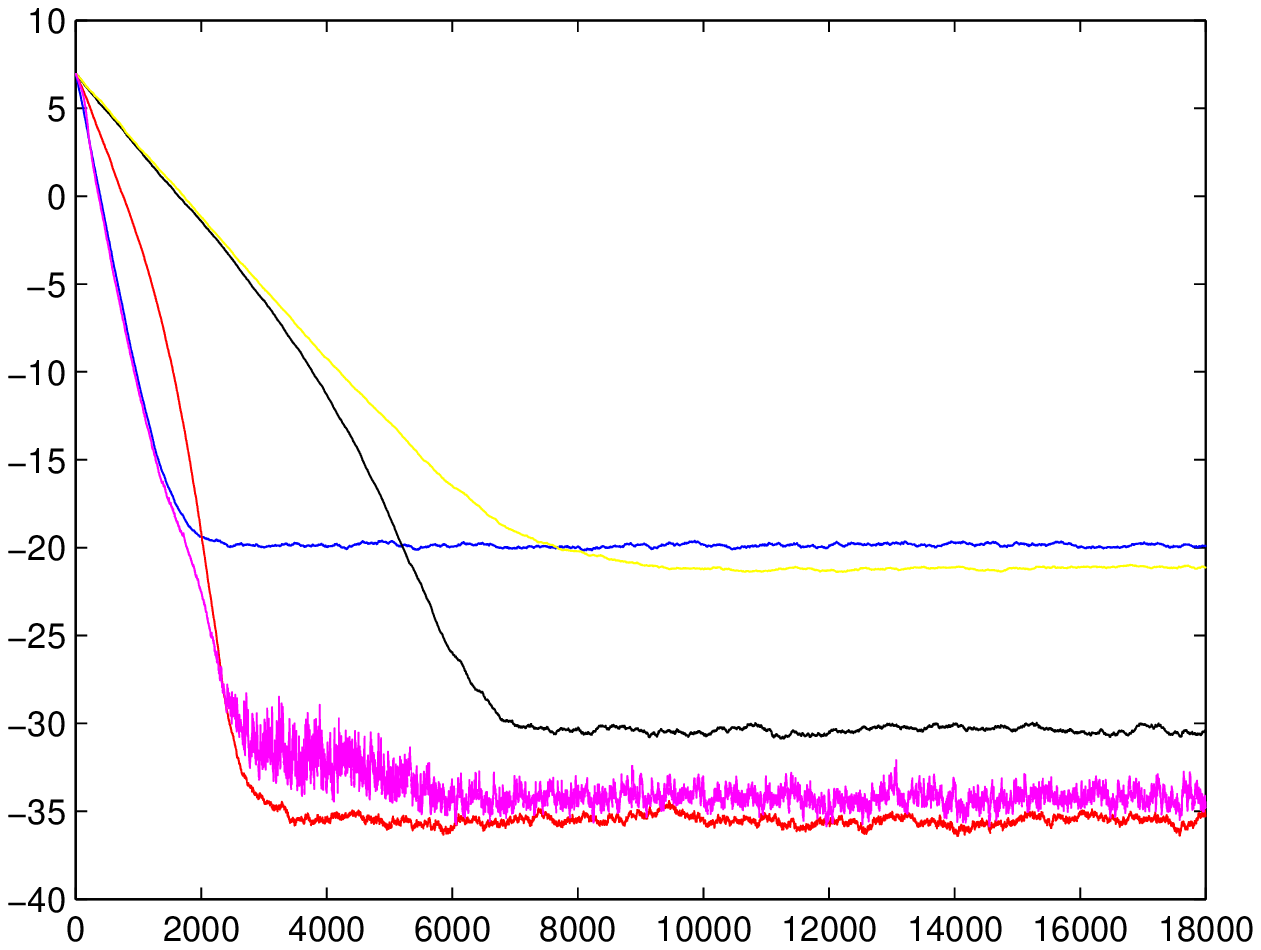}
\end{center}
\caption{MSD vs. Iteration Index for Combination of $M=4$ Adaptive Filters for SNR = $20$dB}.
\vspace*{-3pt}
\end{figure*}

\begin{figure*}[h]
\begin{center}
\includegraphics[width=160mm, height=65mm]{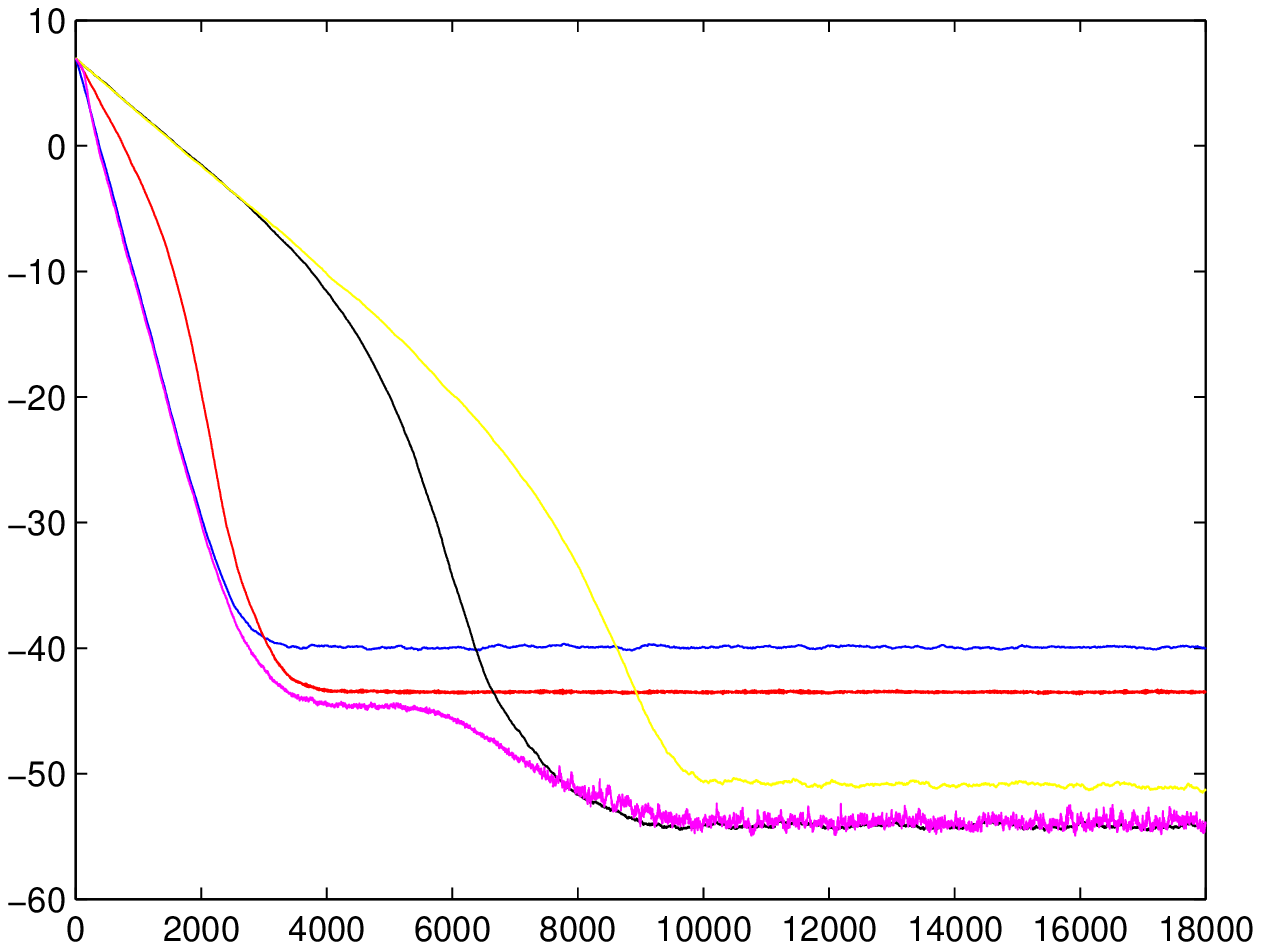}
\end{center}
\caption{MSD vs. Iteration Index for Combination of $M=4$ Adaptive Filters for SNR = $40$dB}.
\vspace*{-3pt}
\end{figure*}

\begin{figure*}[h]
\begin{center}
\includegraphics[width=160mm, height=65mm]{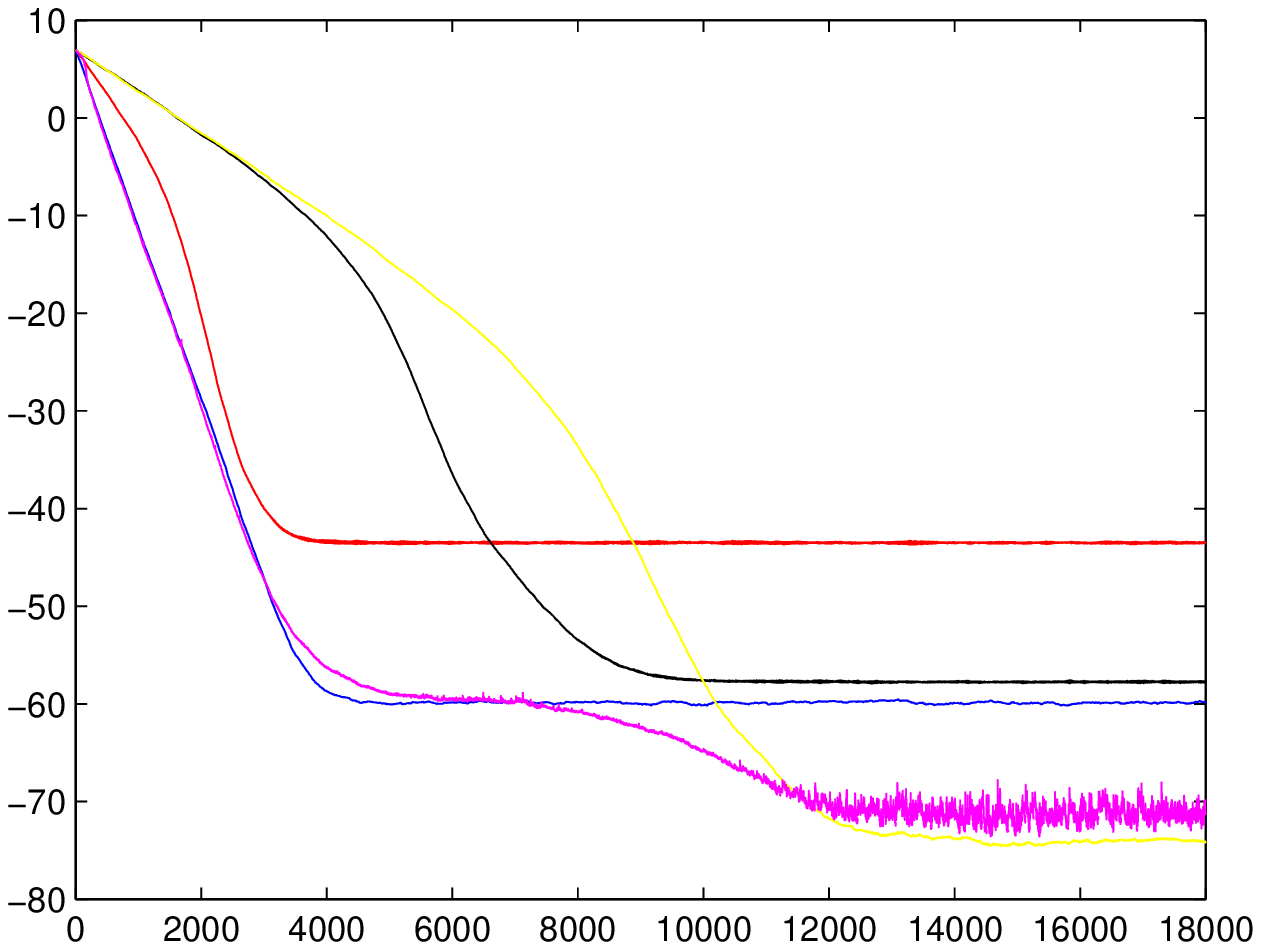}
\end{center}
\caption{MSD vs. Iteration Index for Combination of $M=4$ Adaptive Filters for SNR = $60$dB}.
\vspace*{-3pt}
\end{figure*}

\begin{figure*}[h]
\begin{center}
\includegraphics[width=160mm, height=65mm]{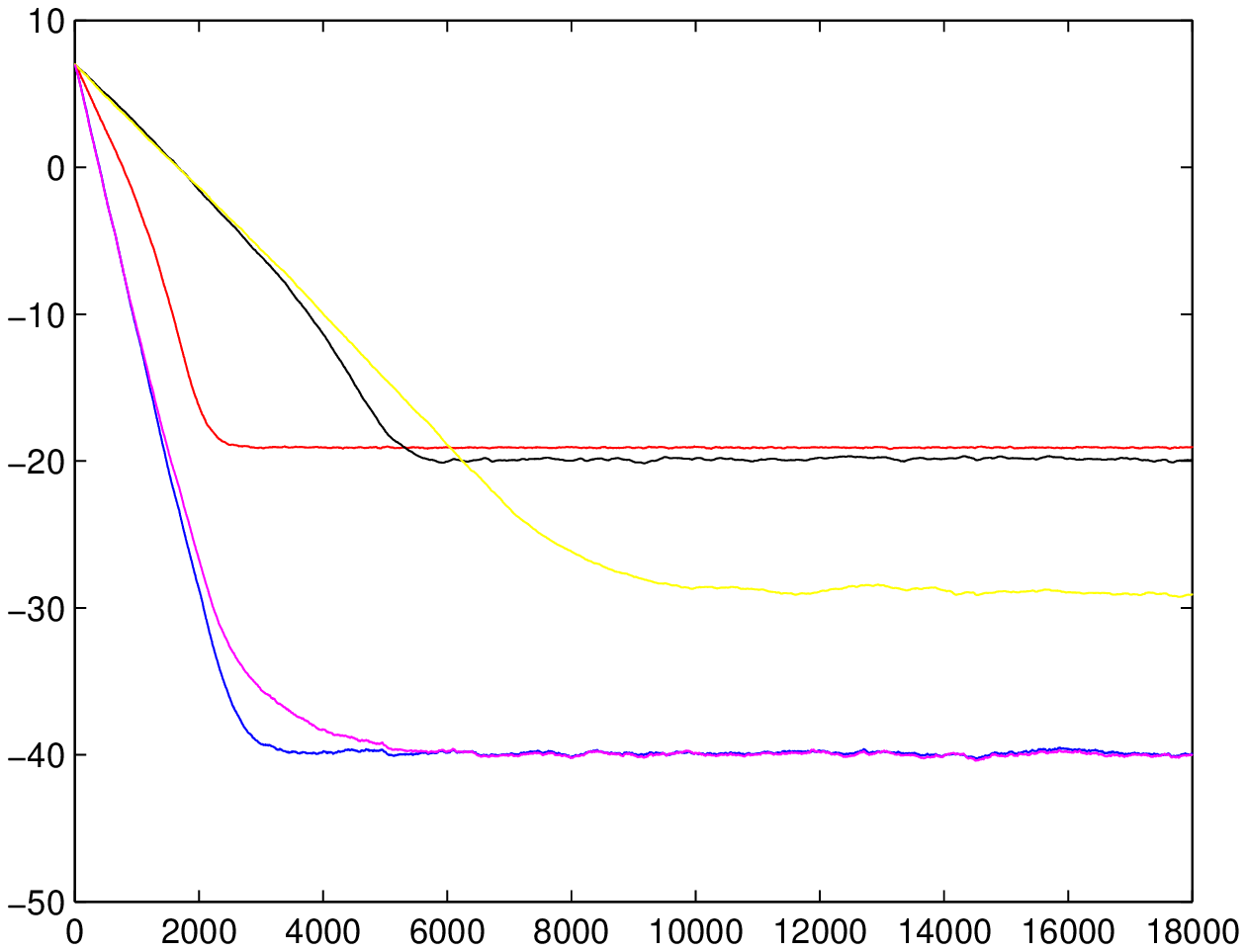}
\end{center}
\caption{MSD vs. Iteration Index for Combination of $M=4$ Adaptive Filters for SNR = $40$dB and the special ${\bf w}_{opt}$}.
\vspace*{-3pt}
\end{figure*}

%
%


\end{document}